\documentclass[aps,twocolumn,showpacs,floatfix]{revtex4-1}
\usepackage{graphicx}
\usepackage{dcolumn}
\usepackage{bm}
\usepackage{epsf}
\usepackage{subfigure}
\usepackage{epstopdf}
\usepackage{amsmath}
\usepackage{amssymb}

\begin{document}
\title{The Three Faces of the Second Law: II. Fokker-Planck Formulation}
\author{Christian Van den Broeck}
\affiliation{Faculty of Sciences, Hasselt University, B-3590 Diepenbeek, Belgium.}
\author{Massimiliano Esposito}
\affiliation{Center for Nonlinear Phenomena and Complex Systems, Universit\'e Libre de Bruxelles, CP 231, Campus Plaine, B-1050 Brussels, Belgium.}

\date{\today}

\begin{abstract}
The total entropy production is the sum of two contributions, the so-called adiabatic and nonadiabatic entropy production, each of which is non-negative. We derive their explicit expressions for continuous Markov processes, discuss their properties and illustrate their behavior on two exactly solvable models.
\end{abstract}



\maketitle
\section{Introduction}\label{Intro}

The second law stipulates that the entropy of an isolated system cannot decrease. In a recent development, \cite{EspositoVdBPRL10}, see also \cite{EspositoHarbola07PRE, ChernyakJarzynski06, Harris07, Ge09}, it was suggested that the second law can in fact be split in two. The total entropy production (EP) $\dot{S}_{tot}$ is the sum of two constitutive parts, namely a so-called adiabatic $\dot{S}_{a}$ and nonadiabatic $\dot{S}_{na}$ contribution. These separate contributions arise from the fact that there are two mechanisms that lead to the time-symmetry breaking characteristic of a dissipative process, namely the application of steady nonequilibrium constraints (adiabatic contribution) or the presence of driving (nonadiabatic contribution). The crucial point is to note that each of these contributions is separately non-negative. We can thus identify ``three faces" to the second law, the positive rate of production of the total EP, of the adiabatic EP and of the nonadiabatic EP:
\begin{eqnarray}
\dot{S}_{tot} \ge 0 \ \ \;, \ \ \dot{S}_{na} \ge 0 \ \ \;, \ \ \dot{S}_{a}  \ge 0. \label{3faces}
\end{eqnarray}
Explicit expressions for $\dot{S}_{tot}$, $\dot{S}_{a}$ and $\dot{S}_{na}$, including a detailed mathematical and physical discussion, were given at the level of a master equation description in the preceding companion paper \cite{EspoVdB10_Da}. However, in many applications, a description on the basis of a Langevin or Fokker Planck equation is more appropriate. The purpose of this paper is to provide a detailed discussion of the adiabatic $\dot{S}_{a}$ and nonadiabatic $\dot{S}_{na}$ EP for such a description. It should be noted that the master equation description is the more general one, including the Langevin and Fokker-Planck case as a special limit. The rather technical transition between both descriptions based on such a limiting procedure is given in the appendix of this paper. In order to be both self-contained and physically motivated, we derive the same expressions for the adiabatic and nonadiabatic EP directly from the Fokker-Planck equation itself in the main text. The application to Langevin equations driven by Gaussian white noise is immediate since there is a mathematical equivalence between Fokker-Planck and Langevin descriptions \cite{Risken}. Note finally that we focus here on the EP rates for which we provide explicit expressions. We have previously obtained the results for the time-integrated trajectory-dependent versions of the various EP contributions \cite{EspositoVdBPRL10}. The latter can be expressed in terms of relative entropies between probabilities for paths in a direct and various types of reverse experiments. While these results have a profound meaning by revealing the temporal symmetry breaking associated to each contribution, the expressions for the EP rates given here do not refer to any reverse experiment and are thus much easier to measure or calculate.

\section{Fokker Planck equation}

\subsection{Total entropy balance}

Our starting point is the Fokker-Planck equation describing the time evolution of the probability density ${p}_t={p}_t(x)$ for the variables $x$:
\begin{eqnarray}
\dot{p}_t &=& -  \partial_x J_t, \label{FP} 
\end{eqnarray}
with
\begin{eqnarray}
&& J_t = \sum_{\nu} J^{(\nu)}_t= u_t p_t - D_t \partial_x p_t \label{Fluxtot} \\
&& J^{(\nu)}_t = u^{(\nu)}_t  p_t - D^{(\nu)}_t \partial_x p_t \label{FluxFP}\\
&& u_t = \sum_{\nu} u^{(\nu)}_t  \;\;\;\;\;\; D_t=\sum_{\nu} D^{(\nu)}_t. \label{MobDiff}
\end{eqnarray}
The quantities $p_t,J_t,J^{(\nu)}_t,u^{(\nu)}_t,D^{(\nu)}_t $, etc. are all functions of the state $x$ of the system, although this is not written explicitly for simplicity of notation. The probability density changes in time due to different processes $\nu$. As a result a probability flux $J^{(\nu)}_t$ is associated to each process. We also assume general time-dependent drift and diffusion coefficients $u^{(\nu)}_t$ and $D^{(\nu)}_t$.

We now proceed with the identification of various EP contributions, whose positivity can be guaranteed on a purely mathematical basis. As we will see in the next section, these terms can be interpreted as genuine forms of EP when the proper physical content is taken into account. We focus on the time evolution of the Shannon entropy for the system:
\begin{eqnarray}
S(t) = - \int dx \; p_t \ln p_t \label{Sentropy}.
\end{eqnarray}
Note that we assume here and henceforth that the Boltzmann constant $k_B$ is set equal to 1. Using (\ref{FP}) one finds:
\begin{eqnarray}
\dot{S}(t) &=& - \int dx \; \dot{p}_t \ln p_t \nonumber\\
&=& - \sum_{\nu} \int dx \;  J^{(\nu)}_t \frac{\partial_x p_t}{p_t} \nonumber\\
&=& \sum_{\nu} \int dx \;  J^{(\nu)}_t \bigg( \frac{J^{(\nu)}_t }{D^{(\nu)}_t  p_t } - \frac{u_t}{D^{(\nu)}_t  } \bigg) \label{FPentropy}
\end{eqnarray}
We used partial integration (assuming that the boundary contributions cancel) and the definition of the probability flux (\ref{FluxFP}) to get the results on the second and third line. One identifies the following two parts:
\begin{eqnarray}
\dot{S}(t) &=& \dot{S}_e(t) + \dot{S}_i(t) \label{sl1} .
\end{eqnarray}
The first term corresponds to the entropy flow into the system:
\begin{eqnarray}
\dot{S}_e(t) = - \sum_{\nu} \int dx \; J^{(\nu)}_t \frac{u^{(\nu)}_t}{D^{(\nu)}_t}.\label{FPrEP}
\end{eqnarray}
The second term is the non-negative irreversible EP:
\begin{eqnarray}
\dot{S}_i(t)=\sum_{\nu} \int dx \frac{\big( J^{(\nu)}_t \big)^2}{D^{(\nu)}_t p_t} \geq 0.  \label{FPtotEP}
\end{eqnarray}
These expressions are in agreement with the results given previously in the literature for the case of a single process $\nu$ \cite{Qian01PRE,Seifert05,Jayannavar09,Peliti06}.

We mention two additional properties of the total EP. First, we note the following inequality:
\begin{eqnarray}
\dot{S}_{i}(t) \geq \int dx \frac{\big( J_t \big)^2}{D_t p_t} \geq 0,  \label{FPtotEPCG}
\end{eqnarray}
which shows that the total EP is underestimated if the constituent processes $\nu$ are not properly identified.
The above result follows from the following inequality, valid for any set of numbers $y_i\geq 0$:
\begin{eqnarray}
\sum_i \frac{x^2_i}{y_i} \geq \frac{\big(\sum_i {x_i}\big)^2}{\sum_i {y_i}}.  \label{sumineq}
\end{eqnarray}
To proof this inequality, consider first the case with all $x_i \geq 0$. The above inequality is identical to Jensen's inequality $\langle x/y \rangle \langle y/x \rangle \geq 1$, where the averages are over the variables $x_i/y_i$ and $y_i/x_i$ with respect to the probability distribution $p_i=x_i/\sum_i {x_i}$.  The above inequality will hold a fortiori if not all $x_i \geq 0$, since the l.h.s. is insensitive to a change of sign of the variables $x_i$, while the r.h.s. can only become smaller.

Second, the irreversible total EP has the familiar form of a sum over fluxes times forces:
\begin{eqnarray}
\dot{S}_{i}(t) =\sum_{\nu} \int dx J^{(\nu)}_t X^{(\nu)}_t,
\end{eqnarray}
with the force associated to process $\nu$ given by:
\begin{eqnarray}
X^{(\nu)}_t=X^{(\nu)}_t(x) = \frac{J^{(\nu)}_t(x)}{D^{(\nu)}_t(x) p_t(x)} . \label{ForceFP}
\end{eqnarray}

\subsection{Thermodynamic interpretation}\label{ThermoInt}

The above description can be postulated on a purely phenomenological or mathematical basis, as it corresponds to the general equation of evolution for a continuous Markov process. We now make a number of comments that validate the model and the derived expressions for the various types of EP, from a physical point of view.

A Markovian stochastic evolution for the degree of freedom of interest, the system, originates from the elimination of fast degrees of freedom, the reservoir, that do not need to be described because they are at instantaneous equilibrium for any given state of the relevant variables. Furthermore, it is assumed that the different processes $\nu$ corresponding to groups of fast eliminated variables (reservoirs) at different equilibrium values, do not directly interfere which each other. Otherwise this interaction would need to be described and the set of variables describing completely the physical process would need to be enlarged. The correct identification of the reservoirs $\nu$ is a crucial step since we have seen in (\ref{FPtotEPCG}) that failure to do so will typically lead to an underestimation of the EP. At constant (in time) drift and diffusion coefficient, when all reservoirs but one are switched off, say $\nu$, the system should reach an equilibrium steady state distribution corresponding to the thermodynamic properties of the reservoir $\nu$ and satisfying the condition of detailed balance $X^{(\nu)}_t=0$. In presence of different reservoirs $\nu$, the steady state is out of equilibrium because it will break detailed balance. Indeed, all the reservoirs try unsuccessfully to impose their equilibrium value on the system. The essential step to connect the stochastic description to the present thermodynamics discussion is the local detail balance condition:
\begin{eqnarray}
\frac{u^{(\nu)}_t}{D^{(\nu)}_t} = \beta^{(\nu)} F_t . \label{DBFP} 
\end{eqnarray}
This relation implies to identify the energy $E_t$ of the system, because the force $F_t$ is the negative derivative of the energy: $F_t = - \partial_x E_t$ (for more details, see also appendix and Eq.(22) in \cite{EspoVdB10_Da} for the corresponding relation for the master equation). It also mathematically guaranties that in presence of a single reservoir $\nu$ and of a time-independent external force $F$, the probability distribution of the Fokker-Plank equation (\ref{FP}) will eventually reach the equilibrium distribution $p^{eq} \sim \exp{\{-\beta E\}}$. 
Since (\ref{DBFP}) translates the fact that each of the reservoirs remains at equilibrium, the EP (\ref{sumineq}) is also the total EP since no irreversible processes take place in the reservoirs: $\dot{S}_{i}(t)=\dot{S}_{tot}(t)$. For the same reason, the entropy flow (\ref{FPrEP}) into the system corresponds to minus the entropy change into the reservoirs: $\dot{S}_{e}(t)=-\dot{S}_{r}(t)$. 

By introducing the generalized mobility $\mu^{(\nu)}_t$ of process $\nu$ as $u^{(\nu)}_t = \mu^{(\nu)}_t F_t$, we see that (\ref{DBFP}) is in fact the generalized fluctuation-dissipation Einstein relation $\mu^{(\nu)}_t = \beta^{(\nu)} D^{(\nu)}_t$. This clarifies the thermodynamic meaning of the force (\ref{ForceFP}) since ${J^{(\nu)}_t}/{p_t}$ can be identified as the local speed $v^{(\nu)}$ of process $\nu$ so that $X^{(\nu)}_t = v^{(\nu)}/(\mu^{(\nu)} T^{(\nu)})$ corresponds to the usual thermodynamic expression of a force, speed divided by mobility, over the temperature. 
Furthermore, using (\ref{DBFP}) in (\ref{FPrEP}), we find that the entropy flow takes the familiar thermodynamic form
\begin{eqnarray}
\dot{S}_e(t) = \sum_{\nu} \beta^{(\nu)} \dot{Q}^{(\nu)}(t), \label{EntFlowBis} 
\end{eqnarray}
where heat flowing into the system is given by
\begin{eqnarray}
\dot{Q}^{(\nu)}(t) = -\int dx \; J^{(\nu)}_t F_t. \label{Heat} 
\end{eqnarray}
Introducing the system energy 
\begin{eqnarray}
E(t) = \int dx \; p_t E_t , \label{Energy} 
\end{eqnarray}
we find (using integration by part and neglecting the boundary terms) that the first law of thermodynamics assumes the familiar form:
\begin{eqnarray}
\dot{E}(t) = \dot{W}(t) + \sum_{\nu} \dot{Q}^{(\nu)}(t) , \label{FirstPrinc} 
\end{eqnarray}
where the work is given by 
\begin{eqnarray}
\dot{W}(t) = \int dx \; p_t \dot{E}_t. \label{Work} 
\end{eqnarray}
We have thus shown that the local detail balance condition (\ref{DBFP}) provides an explicit connection to thermodynamics and justifies the names used for the various entropies in the previous section. \\

We should note however that for systems subjected to nonconservative forces, the local detail balance condition (\ref{DBFP}) will not be satisfied. Even in presence of a single reservoir $\nu$, the steady state will break detailed balance and will thus be a nonequilibrium steady state. 

\subsection{Adiabatic and nonadiabatic entropy balance}

To identify a component related to the relaxation of the system, we introduce the instantaneous steady state solution $p^{st}_t$, being the normalized (supposedly unique) solution of the following equation: 
\begin{eqnarray}
\partial_x J^{st}_t = 0, \label{stFP} 
\end{eqnarray}
with
\begin{eqnarray}
&& J^{st}_t=\sum_{\nu} J^{st(\nu)}_t=u_t p^{st}_t - D_t \partial_x p^{st}_t  \nonumber \\
&& J^{st(\nu)}_t = u^{(\nu)}_t p^{st}_t - D^{(\nu)}_t \partial_x p^{st}_t \label{stFluxFP}.
\end{eqnarray}
It corresponds to the the steady state solution of the Fokker Planck equation (\ref{FP}), if the drift and diffusion coefficients are frozen at their instantaneous values. We can now rewrite the flux $J_t^{(\nu)}$, cf. (\ref{FluxFP}), as follows:
\begin{eqnarray}
&&\frac{J_t^{(\nu)}}{p_t} - \frac{J^{st(\nu)}_t} {p^{st}_t} =-D_t^{(\nu)} \partial_x \bigg( \ln\frac{p_t}{p^{st}_t} \bigg) \label{FluxFPNew}
\end{eqnarray}
As a result, the following expression is identically zero:
\begin{eqnarray}
&&\sum_{\nu} \int dx J^{st(\nu)}_t \frac{p_t}{ D_t^{(\nu)} p^{st}_t} \bigg( \frac{J_t^{(\nu)}}{ p_t} -\frac{J_t^{st(\nu)}}{ p_t^{st}}  \bigg)\nonumber \\
&&=  -\int dx J_t^{st} \partial_x \bigg( \frac{p_t}{p^{st}_t} \bigg) = 0 \label{special1}
\end{eqnarray}
The last step in (\ref{special1}) follows from (\ref{stFP}) by partial integration, assuming that the boundary term is zero (infinite system or system with periodic boundary condition). The separation of total EP (\ref{FPtotEP}) in two contributions that are separately positive is now straightforward. We write the integrand in (\ref{FPtotEP}) as follows:
\begin{eqnarray}
\frac {p_t}{D_t^{(\nu)}}\bigg(\frac{J^{(\nu)}_t }{ p_t } \bigg)^2
= \frac {p_t}{D_t^{(\nu)}}\bigg(\frac{J^{(\nu)}_t }{ p_t }-\frac{J^{st(\nu)}_t }{ p^{st}_t }+\frac{J^{st(\nu)}_t }{ p^{st}_t } \bigg)^2. \label{special2}
\end{eqnarray} 
From (\ref{special1}) and  (\ref{special2}), we conclude that the total EP (\ref{FPtotEP}) can be written as follows:
\begin{eqnarray}
\dot{S}_{tot}(t) = \dot{S}_{na}(t)+ \dot{S}_{a}(t),  \label{splittotEP}
\end{eqnarray}
with the following explicit expressions for the nonadiabatic and adiabatic rate of EP:
\begin{eqnarray}
\dot{S}_{na}(t) &=&  \sum_{\nu}\int dx \; \frac{p_t}{D_t^{(\nu)}} \bigg( \frac{J^{(\nu)}_t}{p_t} - \frac{J_t^{{st(\nu)}}}{p_t^{st}} \bigg)^2 \geq 0  \label{FPnaEP} \\
\dot{S}_{a}(t) &=& \sum_{\nu} \int dx \; \frac{p_t}{D_t^{(\nu)}} \bigg( \frac{J_t^{st(\nu)}}{p_t^{st}} \bigg)^2 \geq 0  \label{FPaEP} 
\end{eqnarray}
These quantities are clearly non-negative. The nonadiabatic EP is zero for an infinitely fast relaxing system being all the time in  the instantaneous steady state. The expression for the adiabatic EP is similar to the total EP, but with the steady state contributions ${J_t^{{st(\nu)}}}/{p_t^{st}}$ rather than the actual ${J_t^{{(\nu)}}}/{p_t}$ singled out. Both expressions can also be obtained as the limits of the corresponding expressions for the nonadiabatic and adiabatic EP for a master equation, cf. appendix \ref{Appendix}. In case of a system in contact with a single reservoir and subjected to a nonconservative force (the steady state break detailed balance), the adiabatic EP is the housekeeping heat divided by the reservoir temperature \cite{Oono98,HatanoSasa01,SpeckSeifert05,Harris07,ChernyakJarzynski06}.

We now make a number of further comments including alternative expressions for the adiabatic and nonadiabatic EP. First, from
\begin{eqnarray}
&&\frac{1}{D_t }\bigg(\frac{J_t}{p_t} - \frac{J^{st}_t} {p^{st}_t}\bigg) = \nonumber\\
&&\frac{1}{D^{(\nu)}_t }\bigg(\frac{J_t^{(\nu)}}{p_t} - \frac{J^{st(\nu)}_t} {p^{st}_t}\bigg)
=- \partial_x \bigg( \ln\frac{p_t}{p^{st}_t} \bigg) \label{FluxtotFPNew}
\end{eqnarray}
it follows that the nonadiabatic EP can be written in terms of compound quantities (obtained by summation over the processes $\nu$) only:
\begin{eqnarray}
\dot{S}_{na}(t) &=& \int dx \; \frac{p_t}{D_t} \bigg( \frac{J_t}{p_t} - \frac{J_t^{st}}{p_t^{st}} \bigg)^2 \geq 0. \label{FPnaEP1}
\end{eqnarray}
We conclude that the nonadiabatic EP is not sensitive to the identification of the various separate processes describing the exchange with different reservoirs. The rationale is that this EP reflects relaxation EP within the system itself. We recall furthermore that the total EP, which is the sum of the adiabatic and nonadiabatic contribution, can only decrease upon coarse-graining the processes $\nu$. Hence, we conclude that the adiabatic EP is underestimated in exactly the same way as the total EP, when the constituting processes $\nu$ are not properly identified. 

Second, we mention the following alternative form of the nonadiabatic EP:
\begin{eqnarray}
\dot{S}_{na}(t) =- \int dx \; {\dot{p}_t} \ln\frac{p_t}{p_t^{st}}, \label{FPnaEP2}
\end{eqnarray}
obtained from (\ref{FPnaEP1}) with (\ref{FluxtotFPNew}). This form is quite convenient for a direct calculation of the nonadiabatic EP when the probability distribution $p_t$ is known explicitly. We also note that for constant in time drift and diffusion coefficients, one has that $\dot{H}=-\dot{S}_{na} \le 0$, where $H= \int dx \; {{p}_t} \ln{p_t}/{p_t^{st}}\ge 0$ is a Lyapunov function. This thus proves the convergence of ${p}_t$ to the (supposedly unique) steady state ${p}^{st}$ \cite{KampenB97}.

Third, we introduce another entropic contribution, the so-called excess heat \cite{Oono98}:
\begin{eqnarray}
\dot{S}_{ex}(t) = - \int dx \; J_t \partial_x \ln p^{st}_t. \label{FPexEP}
\end{eqnarray}
This expression allows to complement the familiar EP balance equation (\ref{sl1}) with  two other balance equations, leading to an alternative presentation of the ``three faces" of the second law. Indeed, one immediately verifies that 
\begin{eqnarray}
\dot{S}(t) &=& -\dot{S}_{ex}(t) + \dot{S}_{na}(t) \label{newSecondLawI} \\
\dot{S}_r(t) &=& \dot{S}_{ex}(t) + \dot{S}_{a}(t) \label{newSecondLawII}.
\end{eqnarray}
Each of these balance equations features the sum of an exchange term, the excess entropy, with has no definite sign, plus an irreversible non-negative EP term. The nonadiabatic term is related to the system properties and is independent from the constituting processes $\nu$. This is not the case of the adiabatic term which represents the dissipation incurred via the contacts with the various reservoirs. We note that when considering transitions between steady states, (\ref{newSecondLawI}) becomes the second law of steady state thermodynamics \cite{Oono98,HatanoSasa01}.

Fourth, upon introducing the following thermodynamic forces:
\begin{eqnarray}
&&\hspace{1cm} X^{(\nu)}_t = \frac{J^{(\nu)}_t}{D^{(\nu)}_t p_t} = A^{(\nu)}_t + N_t \label{Xfp} \\
&& A^{(\nu)}_t = \frac{J^{st(\nu)}_t}{D^{(\nu)}_t p^{st}_t} \  \  \;, \  \ 
N_t = \frac{\partial_x p^{st}_t}{p^{st}_t} - \frac{\partial_x p_t}{p_t}, \label{Xfp2}
\end{eqnarray}
each of the irreversible EP terms can be written under the familiar form of a sum over fluxes times forces:
\begin{eqnarray}
\dot{S}_{a}(t) &=& \sum_{\nu} \int dx \; J^{(\nu)}_t A^{(\nu)}_t \label{FPaEP2}\\
\dot{S}_{na}(t) &=& \int dx \; J_t N_t \label{FPnaEP3}
\end{eqnarray}

\section{Applications}

\subsection{Brownian particle in contact with two thermal reservoirs}

We consider an underdamped Brownian particle in contact with two separate heat baths at temperature $T^{(1)}$ and $T^{(2)}$. Such a model has been studied in the context of an analysis of the Feynmann ratchet \cite{ParrondoEspagnol96}. It corresponds to the simplest model for thermal conduction by a single degree of freedom. The more suggestive presentation is via an equation of motion written under the form of a Langevin equation:
\begin{eqnarray}
\dot{v} = -(\gamma^{(1)}_t+\gamma^{(2)}_t) v +\sqrt{2\gamma^{(1)}_t T^{(1)}} \xi^{(1)}+\sqrt{2\gamma^{(2)}_t T^{(2)}} \xi^{(2)}, \nonumber \\ \label{FPvit}
\end{eqnarray}
with $\xi^{(1)}$ and $\xi^{(2)}$ independent Gaussian white noises of intensity $1$. The variable ${v}$ plays the role of the "speed" of the Brownian particle (although such a speed does not exist in a mathematical sense, only its increments are well defined), while $\gamma^{(1)}_t$ and $\gamma^{(2)}_t$ are the friction coefficients appearing due to the contact with the respective reservoirs $1$ and $2$ which we assume externally controllable (even if this might be physically not very realistic it serves to illustate our results).  The mass of the particle is taken equal to unity. Note that we do not take into account a spatial degree of freedom. This further simplification corresponds to a thermal contact tightly bound to a specific location (in a ``delta function" potential). 

In the context of the Fokker Planck description, we identify the following two drift and diffusion coefficients ($\nu=1,2$):
\begin{eqnarray}
u^{(\nu)}_t =-\gamma^{(\nu)}_t v \  \ \;, \  \  D^{(\nu)}_t= \gamma^{(\nu)}_t T^{(\nu)}.
\end{eqnarray}
Note that we have incorporated the appropriate fluctuation dissipation theorem through the relation linking the diffusion to the friction coefficient (here, $E=v^2/2$ corresponds to the kinetic energy). We mention a further peculiarity of this model. The Langevin equation can be rewritten as:
\begin{eqnarray}
\dot{v} = -\gamma_t v +\sqrt{2 D_t} \xi
\end{eqnarray}
with
\begin{eqnarray}
&&u_t=\sum_{\nu} u^{(\nu)} =-\gamma_t v \nonumber\\
&& \gamma_t = \sum_{\nu}  \gamma^{\nu}_t \ \;\ \ D_t=\sum_{\nu} D^{(\nu)}_t=\gamma_t T_t.
\end{eqnarray}
Consequently, if the distinction between the two processes $\nu$ is not made, this situation corresponds to Brownian particle in contact with a single heat bath at temperature
\begin{eqnarray}
T_t = \frac{\gamma^{(1)}_t T^{(1)} + \gamma^{(2)}_t T^{(2)}}{\gamma^{(1)}_t+\gamma^{(2)}_t}
\end{eqnarray}
and the steady state corresponds to equilibrium (zero EP, with equipartition $\langle{v^2}\rangle_t = T_t$, see also below). This point illustrates our discussion concerning the physical input needed to validate the expression for EP, and in particular the correct identification of the basic processes that are taking place. As mentioned before, the nonadiabatic EP will be correctly reproduced, but both total and adiabatic EPs will be underestimated. In the present case, the underestimation is dramatic since the stationary distribution of the reduced description corresponds to thermal equilibrium, so that the coarse-grained adiabatic EP will be identically zero.

The probability distribution $p_t(v)$ for the speed $v$ obeys  the following Fokker-Plank equation:
\begin{eqnarray}
\dot{p}_t(v) = \gamma_t \partial_{v}[   v p_t(v)+  T_t \partial_{v} p_t(v)]. \label{MotionGF}
\end{eqnarray}
The solution to this equation is a Gaussian distribution, if it is so initially:
\begin{eqnarray}
p_t(v)=\frac{1}{\sqrt{2\pi \langle v^2 \rangle_t}} \exp{\bigg(-\frac{(v-\langle v \rangle_t)^2}{2 \langle v^2 \rangle_t}\bigg)}. \label{GaussSol}
\end{eqnarray}
Its time evolution is completely determined by that of the first and second moments, obeying the following set of equations:
\begin{eqnarray}
\partial_t \langle v \rangle_t &=& - \gamma_t \langle v \rangle_t   \label{smbis} \\
\partial_t \langle v^2 \rangle_t &=& -2 \gamma_t (\langle v^2 \rangle_t - T_t) \label{sm}.
\end{eqnarray}

The steady state form is given by  
\begin{eqnarray}
p^{st}_t(v)=\frac{1}{\sqrt{2\pi T_t}} \exp{\bigg(-\frac{v^2}{2T_t}\bigg)}.
\end{eqnarray}
One easily verifies that: 
\begin{eqnarray}
&&J^{(\nu)}_t=  \frac{Q^{(\nu)}(t)}{\langle v^2 \rangle_t}  p_t v \ \ , \ \ Q^{(\nu)}(t)= \gamma^{(\nu)} \big( T^{(\nu)} - \langle v^2 \rangle_t \big) \nonumber \\
&&A^{(\nu)}_t= \big(\frac{1}{T_t}-\frac{1}{T^{(\nu)}}\big) v \ \ , \ \  N_t= \big(\frac{1}{\langle v^2 \rangle_t}-\frac{1}{T_t} \big) v.
\end{eqnarray}
It is also convenient to define $Q(t)=Q^{(1)}(t)+Q^{(2)}(t)$.
Inserting the above results in the expression for the various forms of irreversible EP, we find:
\begin{eqnarray}
&&\dot{S}_{a}(t) = \frac{Q(t)}{T_t} - \sum_{\nu} \frac{Q^{(\nu)}(t)}{T^{(\nu)}} 
= \frac{\gamma_1 \gamma_2}{\gamma} \frac{\langle v^2 \rangle_t}{T_t} \frac{( T_1-T_2 )^2}{T_1T_2} \nonumber \\
&&\dot{S}_{na}(t) = Q(t) \big(\frac{1}{\langle v^2 \rangle_t}-\frac{1}{T} \big) = \frac{Q^2(t)}{\gamma T \langle v^2 \rangle_t} \nonumber \\
&&\dot{S}_{tot}(t)=\dot{S}_{a}(t)+\dot{S}_{na}(t)\label{EPmodel1}
\end{eqnarray}
For the initial confition $p_{t=0}(v)=\delta(v)$ and in absence of externally control of the friction coefficients, the equation for the second moment reads:
\begin{eqnarray}
\langle{v^2}\rangle_t = T (1- e^{-2 \gamma t}), \label{sms}
\end{eqnarray}
which lead to the simplification: 
\begin{eqnarray}
&&\dot{S}_{a}(t) = \frac{\gamma_1 \gamma_2}{\gamma} (1-e^{-2\gamma t}) \frac{( T_1-T_2 )^2}{T_1T_2} \nonumber\\
&&\dot{S}_{na}(t) =\gamma \frac{e^{-4\gamma t}}{1-e^{-2\gamma t}} \nonumber\\
&&\dot{S}_{tot}(t)=\dot{S}_{a}(t)+\dot{S}_{na}(t) \label{EPmodel1bis}
\end{eqnarray}
These EPs reproduce expected properties. In absence of external driving the nonadiabatic EP decays to zero as the steady state (where $\langle v^2 \rangle=T$) is approached. The adiabatic contribution associated to the application of the non-equilibrium boundary conditions tends toward the usual thermodynamic expression for the EP associated to a steady heat flux between two reservoirs by a device with a thermal conductivity equal to $\kappa={\gamma_1 \gamma_2}/{\gamma}$ \cite{ParrondoEspagnol96}. We finally note that Fourier law is recovered in the steady state $\dot{Q}_1=-\dot{Q}_2=\kappa (T_1-T_2)$. 

\subsection{Driven Brownian particle on a circle}

\begin{figure}[h]
\centering
\begin{tabular}{c@{\hspace{0.5cm}}c}
\hspace{0.35cm}
\rotatebox{0}{\scalebox{0.35}{\includegraphics{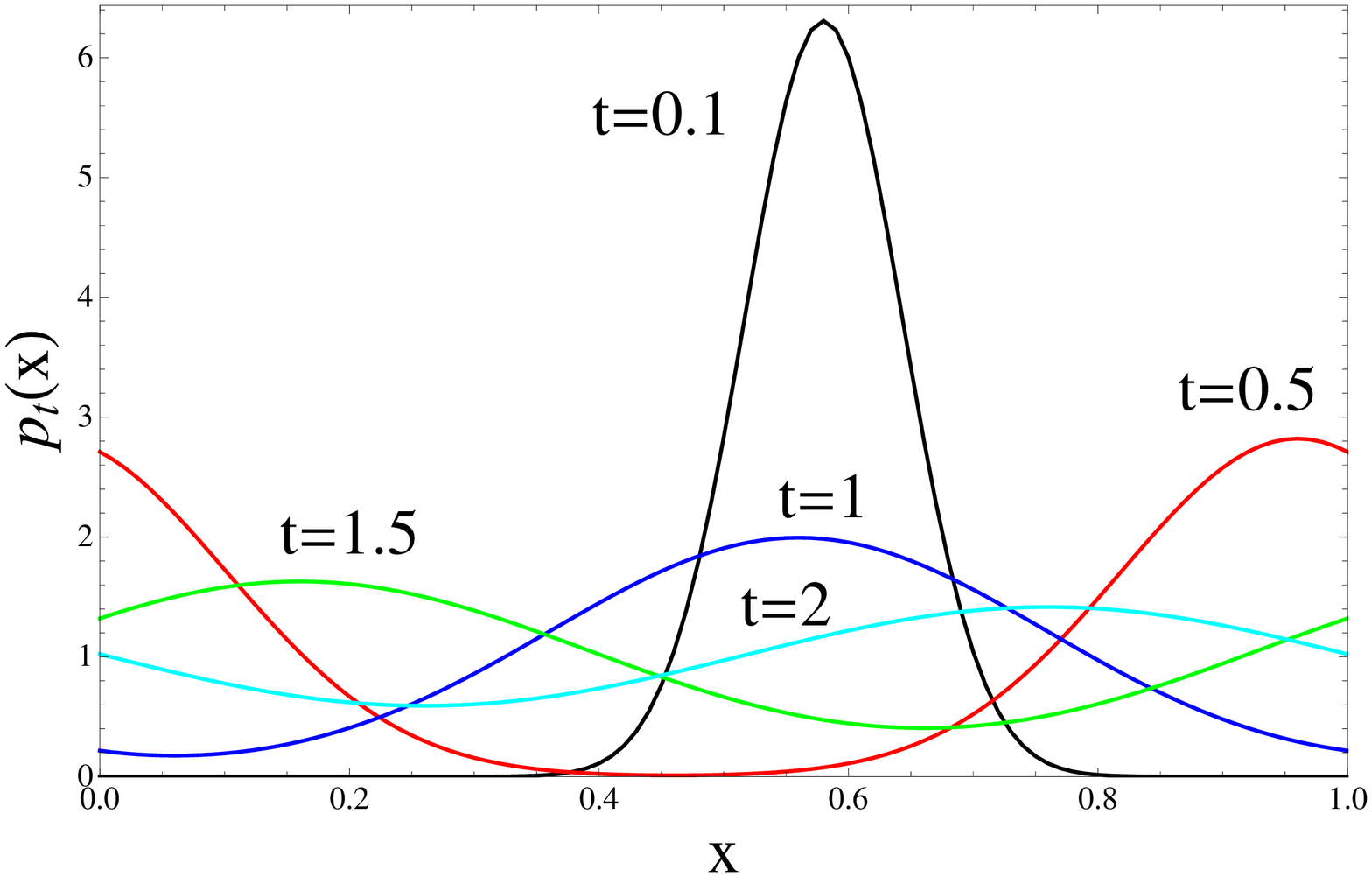}}} \vspace{0.35cm}\\ \hspace{0.5cm}
\rotatebox{0}{\scalebox{0.34}{\includegraphics{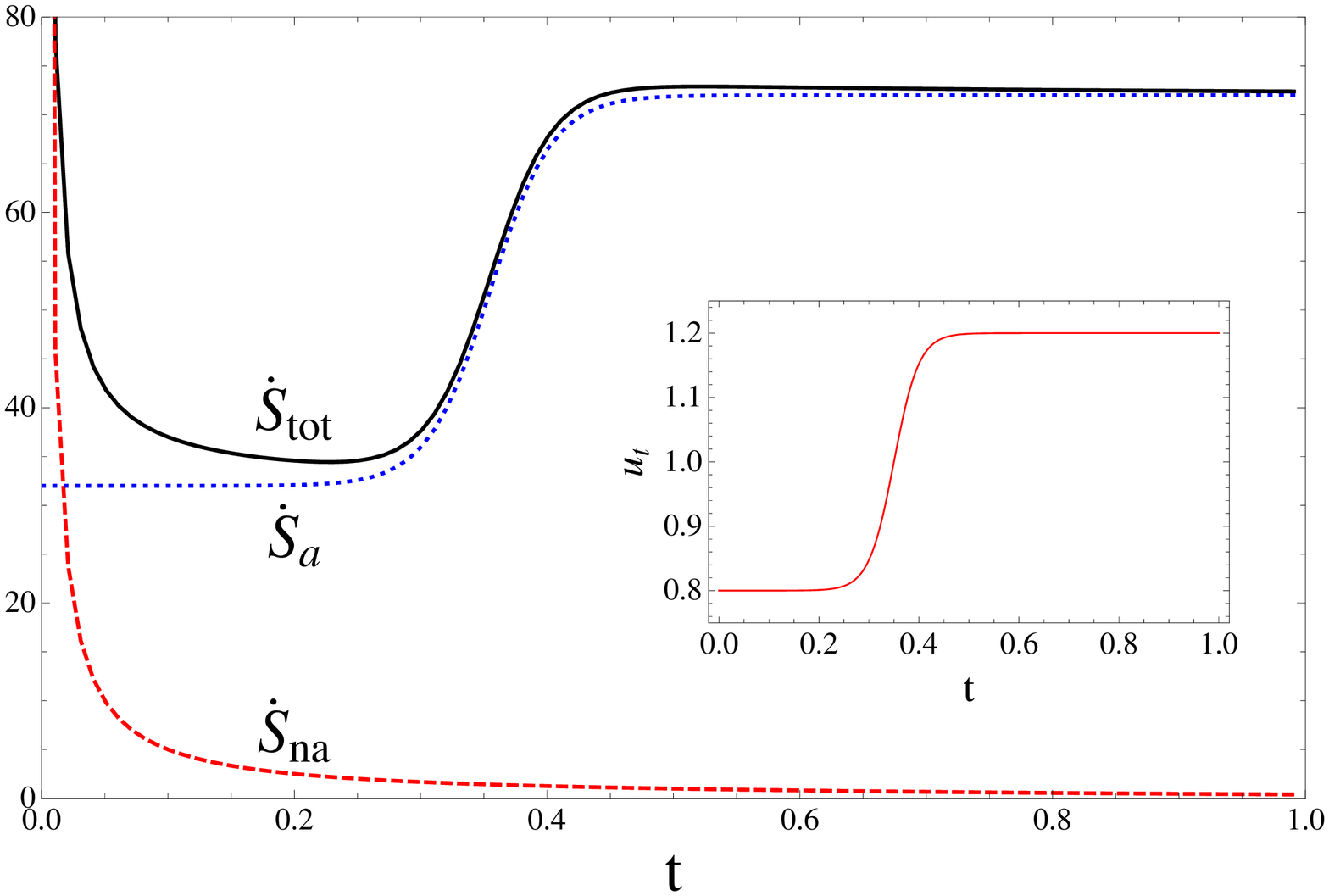}}}
\end{tabular}
\caption{(Color online) Probability distribution at different times and total, nonadiabatic and adiabatic EP ($D=0.02$). The initial divergence of the nonadiabatic EP is due to the singular initial condition $p_{t=0}=\delta(x)$. As the uniform distribution is approached, the nonadiabatic EP decreases. The application of a switch from the initial value 0.8 to the value 1.2 of the drift, cf.  inset, has no effect on the nonadiabatic EP, but results in an additional adiabatic EP. }
\label{plot}
\end{figure}

We next consider an overdamped Brownian particle
\begin{eqnarray}
\dot{x} = u_t +\sqrt{2 D} \xi,   \label{FPex2EP}
\end{eqnarray}
where $u_t$ and $D$ are the (time dependent) drift  and (time independent) diffusion coefficient  both position independent. We furthermore assume $x \in [0,1]$ with periodic boundary conditions. This model was discussed in stochastic thermodynamics as a simple example of  a particle driven by a nonconservative force \cite{Seifert06}. For simplicity, we consider the special initial condition $p_{t=0}=\delta(x)$. The exact time-dependent solution of the Fokker Planck equation is expressed in terms of the well-known solution on the infinite line:
\begin{eqnarray}
p^{*}_t(x)=\frac{1}{\sqrt{4\pi D t }} e^{-\frac{(x-\int_0^t d\tau u(\tau) )^2}{4Dt}},
\end{eqnarray}
namely ($x \in [0,1]$):
\begin{eqnarray}
p_t(x)=\sum\limits_{n=-\infty}^{+\infty} p^{*}_t(x+n).\label{espt}
\end{eqnarray}
It converges to the steady state solution:
\begin{eqnarray}
p^{st}_t(x) = 1.
\end{eqnarray}
This model has a peculiarity: the steady state distribution is identical to the equilibrium distribution $p^{st}_t(x)=p^{eq}(x)=1$. 
As a consequence one has that $J^{st}_t=u_t$. Furthermore, once $p_t$ has relaxed to this distribution, this remains  the case even while a  time-dependent driving $u_t$ is still applied. 
We find that the various EPs read:
\begin{eqnarray}
&&\dot{S}_{a}(t) = \frac{u_t^2}{D} \  \ , \  \ \dot{S}_{na}(t) =  D \int_{0}^{1} dx \; \frac{[\partial_x {p_t(x)}]^2 }{p_t(x)}   \nonumber\\
&&\dot{S}_{tot}(t)=\dot{S}_{a}(t)+\dot{S}_{na}(t) \label{EPmodel2}
\end{eqnarray}
The adiabatic EP is proportional to the square of the externally applied drift. The nonadiabatic EP is given by a more complicated expression, but goes to zero as the probability distribution relaxes to the uniform stationary distribution. It remains zero once this distribution is reached, even if a time-dependent driving $u_t$ persists. As an illustration, we reproduce the results for the various EP contributions in Fig. (\ref{plot}), including the effect of a switch in the driving speed. 

\section{Conclusion}\label{conc}

In this paper, we have identified the non-negative EP as well as its two non-zero contributions, the adiabatic and the nonadiabatic part, for Fokker-Planck dynamics. This parallels a similar identification  for master equation dynamics presented in the companion paper \cite{EspoVdB10_Da}. We have shown that this identification allows to ``split the second law in two parts". It remains to be seen what are the implications of this ``doubling" of the second law. In particular, we speculate that it should imply the impossibility of some physical phenomena, being incompatible with the inequalities, that it may provide novel limits, for example on efficiencies of machines, or may be linked to novel symmetries, such as the symmetry of Onsager coefficients. Finally, we reiterate that the positivity of the adiabatic and the nonadiabatic EP obtained here follows by Jensen's inequality from the detailed fluctuation theorems derived in \cite{EspositoVdBPRL10}. The latter deal with the trajectory-dependent adiabatic and nonadiabatic EP, and thus reveal  a much more detailed and deeper statistical symmetry deriving from micro-reversibility. 

\appendix
\section{Fokker-Plank limit of the master equation}\label{Appendix}

In the companion paper \cite{EspoVdB10_Da}, we derive the adiabatic and nonadiabatic EP for Markov processes obeying a master equation (cf. eq.  (7)). The results for the Fokker Planck equation given in the main text can be derived by applying an appropriate limiting procedure, similar to that of Ref. \cite{VdB90}. As a starting point it suffices to consider the case of a tri-diagonal transition matrix, i.e., the only non-zero non-diagonal elements of $W_{m \pm 1,m}^{(\nu)}(\lambda_t)$ are those with $m'=m \pm 1$. The master equation thus has the following form: 
\begin{eqnarray}
\dot{p}_m = - \sum_{\nu}  \big( J_{m+1,m}^{(\nu)}(t) - J_{m,m-1}^{(\nu)}(t) \big), \label{MEqFP_AP}
\end{eqnarray}
where
\begin{eqnarray}
J_{m,m-1}^{(\nu)}(t) = W_{m,m-1}^{(\nu)}(\lambda_t) p_{m-1}(t) - W_{m-1,m}^{(\nu)}(\lambda_t) p_m(t). \nonumber
\end{eqnarray}
We introduce
\begin{eqnarray}
2 D_m^{(\nu)}(\lambda_t) &=& W_{m,m-1}^{(\nu)}(\lambda_t) + W_{m-1,m}^{(\nu)}(\lambda_t) \\
u_m^{(\nu)}(\lambda_t) &=& W_{m,m-1}^{(\nu)}(\lambda_t) - W_{m-1,m}^{(\nu)}(\lambda_t).
\end{eqnarray}
The idea is that the (general) nearest neighbor random walk in the variable $m$ goes over into a (general) diffusion process for a continous variable  $x = m \epsilon$. We illustrate the procedure for $x\in[-L,L]$ with reflecting boundary conditions, covering in the limit $L \rightarrow \infty$ the case of real variables. A similar procedure can be applied for periodic boundary conditions. We consider $m=-N,-(N-1), \cdots0,1,2,\cdots,N$, with reflecting boundary conditions, $W_{N+1,N}=W_{-N-1,-N}=0$. We take the limit $\epsilon \to 0$ and $N \to \infty$, where $N= L/\epsilon$, with $L$ fixed, obtaining a continuous variable $x = m \epsilon \in[-L,L]$. Using $\exp{\{\pm \partial_m\}} f_m = f_{m \pm 1}$ and 
\begin{eqnarray}
&& p_t \equiv p(x,t) = p_m(t) / \epsilon \\
&& \partial_x = \epsilon^{-1} \partial_m \\
&& D^{(\nu)}_t \equiv D^{(\nu)}(x,\lambda_t) = D_m^{(\nu)}(\lambda_t) \epsilon^2 \\
&& u^{(\nu)}_t \equiv u^{(\nu)}(x,\lambda_t) = u_m^{(\nu)}(\lambda_t) \epsilon ,
\end{eqnarray}
we find that (\ref{MEqFP_AP}) goes over into the Fokker-Plank equation:
\begin{eqnarray}
\dot{p}_t &=& - \sum_{\nu} \partial_x J^{(\nu)}_t \label{FP_AP} \\
J^{(\nu)}_t &=& u^{(\nu)}_t p_t - D^{(\nu)}_t \partial_x p_t \label{FluxFP_AP},
\end{eqnarray}
with zero flux boundary conditions.
Similarly the forces (28)-(30) in \cite{EspoVdB10_Da} become
\begin{eqnarray}
&&\hspace{1cm} X^{(\nu)}_t = \frac{J^{(\nu)}_t}{D^{(\nu)}_t p_t} = A^{(\nu)}_t + N_t \label{Xfp_AP} \\
&& A^{(\nu)}_t = \frac{J^{st(\nu)}_t}{D^{(\nu)}_t p^{st}_t} \  \  \;, \  \ 
N_t = \frac{\partial_x p^{st}_t}{p^{st}_t} - \frac{\partial_x p_t}{p_t}. \label{Xfp2_AP}
\end{eqnarray}
Using these results, it is easy to verify that the various EP from section II of \cite{EspoVdB10_Da} lead to the EP of section II of the present paper.

Finally we note that the local detailed balance condition with respect to the various processes $\nu$ given by eq. (22) in \cite{EspoVdB10_Da} reduces in the Fokker-Plank limit to (\ref{DBFP}). 

\section*{Acknowledgements}

M. E. is supported by the Belgian Federal Government (IAP project ``NOSY").

%
\end{document}